# Nonperturbative Studies of Quantum Gravity [1]

Wolfgang Beirl, Harald Markum, Jürgen Riedler

*Institut für Kernphysik, Technische Universität Wien, A-1040 Vienna, Austria*

**Abstract**

We investigate quantum gravity in the path integral formulation using the Regge calculus. Restricting the quadratic link lengths of the originally triangular lattice the path integral can be transformed to the partition function of a spin system with higher couplings on a Kagomé lattice. Various measures acting as external field were considered. Extensions to matter fields and higher dimensions are discussed.

More than thirty years ago Regge proposed to approximate Riemannian manifolds by piecewise flat, simplicial manifolds. He developed a discrete description of General Relativity in which space-time is triangulated by a simplicial lattice, the Regge skeleton [1]. Thus, the lattice becomes a dynamical object, with the edge lengths describing the evolution of space-time. More precisely, dynamics comes from the variation of the link lengths keeping the connectivity of the triangulation fixed. We briefly mention that a somehow complementary method is provided by the dynamical triangulation approach in which the length of the links is kept fixed and the incidence matrix is varied.

Within the Regge scheme the Einstein-Hilbert action for lattice gravity in two dimensions is given by [2]

$$S = \lambda \sum_t A_t - 2k \sum_v \delta_v ,  \qquad (1)$$

with $\lambda$ a cosmological constant and $A_t$ the area of triangle $t$. The curvature of the surface is concentrated at the vertices of the lattice. The deficit angle at a single vertex $v$ is defined by

$$\delta_v = 2\pi - \sum_{t \supset v} [\text{vertex angle at } v] ,  \qquad (2)$$

where $t \supset v$ denotes the triangles meeting at $v$.

The corresponding continuum action is

$$S_c = \int d^2x \sqrt{g}(\lambda - kR) ,  \qquad (3)$$

in which $g$ is the determinant of the metric tensor and $R$ the curvature scalar. According to the Gauss-Bonnet theorem the Einstein part of the action is a topological invariant equal to $4\pi$ times the Euler characteristic of the surface. Therefore, for a manifold with fixed topology the term proportional to $k$ can be dropped. In the following only lattices with the topology of a torus will be considered.

---

[1] Supported in part by FWF under Contract P9522-PHY.

In the path integral formulation a quantization of the above action proceeds by evaluating the expression

$$Z = \int d\mu[q] e^{-S[q]} .\qquad(4)$$

Unfortunately, a unique prescription for the measure does not exist, however, an appropriate choice proposed in the literature is [3]

$$\int d\mu[q] = \prod_l \int \frac{dq_l}{q_l^m} ,\qquad(5)$$

with $m \in \mathbb{R}$ defining a one-parameter family.

The central idea of this investigation is to transform the path integral to a partition function of a spin system. Therefore, we allow all the squared edge lengths to take two values

$$q_l = 1 + \epsilon \sigma_l , \quad 0 \leq \epsilon < 0.6, \quad \sigma_l \in Z_2 .\qquad(6)$$

The real parameter $\epsilon$ is restricted to fulfill the Euclidean triangle inequalities for the $q_l$'s. To rewrite the action in terms of $\sigma_l$ we consider a single triangle (cf. Figure 1).

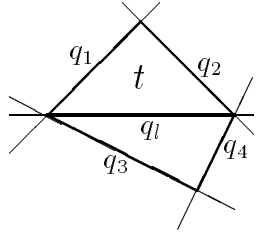

Fig.1: Notation for our triangular lattice. Triangle $t$ consists of edges with the quadratic lengths $q_1$, $q_2$, $q_l$.

Its (squared) area can be expressed by

$$\begin{aligned} A_t^2 &= \begin{vmatrix} q_1 & \frac{1}{2}(q_1 + q_2 - q_l) \\ \frac{1}{2}(q_1 + q_2 - q_l) & q_2 \end{vmatrix} \\ &= \frac{3}{4} + \frac{1}{2}(\sigma_1 + \sigma_2 + \sigma_l)\epsilon + \frac{1}{2}(\sigma_1\sigma_2 + \sigma_1\sigma_l + \sigma_2\sigma_l - \frac{3}{2})\epsilon^2 . \end{aligned}\qquad(7)$$

Expanding $\sqrt{A_t^2}$ the series consists only of terms up to $\sigma^3$ since $\sigma_l^2 = 1$. This suggests the following ansatz

$$A_t = c_0(\epsilon) + c_1(\epsilon)(\sigma_1 + \sigma_2 + \sigma_l) + c_2(\epsilon)(\sigma_1\sigma_2 + \sigma_1\sigma_l + \sigma_2\sigma_l) + c_3(\epsilon)\sigma_1\sigma_2\sigma_l ,\qquad(8)$$

where the coefficients $c_i$ can be obtained by comparing with those in (7) (cf. Figure 2). Using (6) the measure (5) can be replaced by

$$\prod_l \int \frac{dq_l}{q_l^m} \to \sum_{\sigma_l = \pm 1} \exp[-m \sum_l \ln(1 + \epsilon \sigma_l)] ,\qquad(9)$$

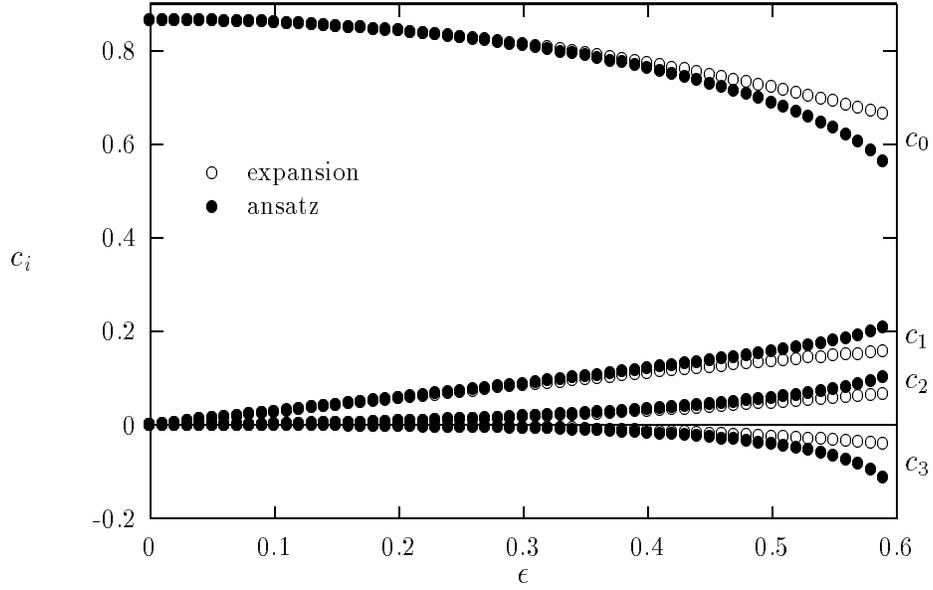

Fig. 2: Coefficients $c_i$ obtained from first terms of series expansion (○) and semianalytically by comparison of coefficients (●).

and the exponential in terms of $\epsilon$ is reduced to

$$m \sum_l \ln(1 + \epsilon \sigma_l) = N_1 m_0(\epsilon) + \sum_l m_1(\epsilon) \sigma_l \ , \qquad (10)$$

with $m_0 = -\frac{1}{2} m \epsilon^2 + O(\epsilon^4)$, $m_1 = m(\epsilon + \frac{1}{3}\epsilon^3) + O(\epsilon^5)$ and $N_1$ the total number of links. Inserting (8) and (9) into the partition function (4) yields

$$\begin{aligned} Z &= \sum_{\sigma_l = \pm 1} J \exp\{-\sum_l (2\lambda c_1 + m_1)\sigma_l - \\ &\quad -\lambda \sum_t [c_2(\sigma_1 \sigma_2 + \sigma_1 \sigma_l + \sigma_2 \sigma_l) + c_3 \sigma_1 \sigma_2 \sigma_l]\} \\ &= \sum_{\sigma_l = \pm 1} J \exp\{-\sum_l [(2\lambda c_1 + m_1)\sigma_l + \\ &\quad + \frac{1}{2}\lambda c_2 (\sigma_1 + \sigma_2 + \sigma_3 + \sigma_4)\sigma_l + \frac{1}{3}\lambda c_3 (\sigma_1 \sigma_2 + \sigma_3 \sigma_4)\sigma_l]\} \ , \end{aligned} \qquad (11)$$

with $J = \exp(-\lambda N_2 c_0 - N_1 m_0)$ and $N_2$ the total number of triangles. Thus, the path integral became the partition function of a system consisting of a spin $\sigma_l$ at each link $l$, with an external "magnetic field" and with two- and three-spin nearest neighbor interactions. Assigning the spin to the corresponding link of the original triangular lattice and drawing the interactions as lines a Kagomé lattice is obtained (cf. Figure 3). Removing the term linear in $\sigma$ by a convenient choice of measure ($m_1 = -2\lambda c_1$) and neglecting three-spin couplings we get the partition function of an Ising model on a Kagomé lattice. To compute the critical coupling we abbreviate $Q = -\frac{1}{2} c_2 \lambda$ and transform the partition function $Z_K(Q)$ of the Kagomé lattice to the partition function $Z_D(L)$ of the decorated honeycomb lattice by applying the star-triangle transformation [4]:

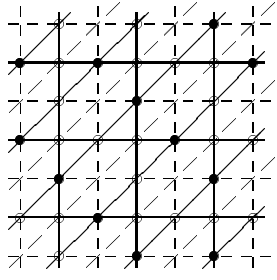

Fig. 3: The triangular lattice is drawn by dashed lines and the Kagomé lattice by solid lines. The spins are represented by ○ and ● characterizing a typical configuration to be discussed in the text.

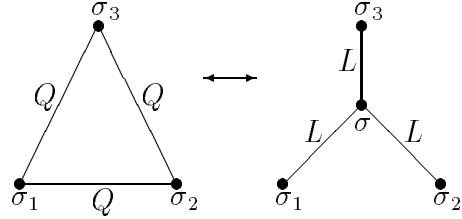

$$\Delta \exp[Q(\sigma_1\sigma_2 + \sigma_2\sigma_3 + \sigma_3\sigma_1)] = \sum_{\sigma=\pm 1} \exp[L\sigma(\sigma_1 + \sigma_2 + \sigma_3)] \ . \tag{12}$$

For $\sigma_1 = \sigma_2 = \sigma_3 = 1$ and $\sigma_1 = \sigma_2 = -\sigma_3 = 1$ we get $2\cosh(3L) = \Delta \exp(3Q)$ and $2\cosh L = \Delta \exp(-Q)$, respectively. Any other choice of $\sigma_i$ coincides with one of these two cases. Hence, $\Delta$ and $Q$ are determined from

$$\Delta^4 = e^{4Q}(e^{4Q} + 3)^2$$
$$e^{4Q} = 2\cosh(2L) - 1 \ . \tag{13}$$

Carrying out this transformation at every vertex gives $Z_D(L) = \Delta^{\frac{2}{3}N_H} Z_K(Q)$ where $N_H$ is the number of vertices of the honeycomb lattice. Furthermore, we can express $Z_D$ by the partition function $Z_H$ of the conventional honeycomb lattice via the decoration-iteration transformation [4]:

$$\sum_{\sigma=\pm 1} \exp[L\sigma(\sigma_1 + \sigma_2)] = I \exp(K\sigma_1\sigma_2) \ . \tag{14}$$

Similarly, we have

$$I = 2e^K$$
$$e^{2K} = \cosh(2L) \ , \tag{15}$$

and inserting (13) and (15) we obtain

$$Z_D(L) = [e^{2Q}(e^{4Q} + 3)]^{\frac{1}{3}N_H} Z_K(Q) = (2e^K)^{N_H} Z_H(K)$$
$$e^{4Q} = 2e^{2K} - 1 \ . \tag{16}$$

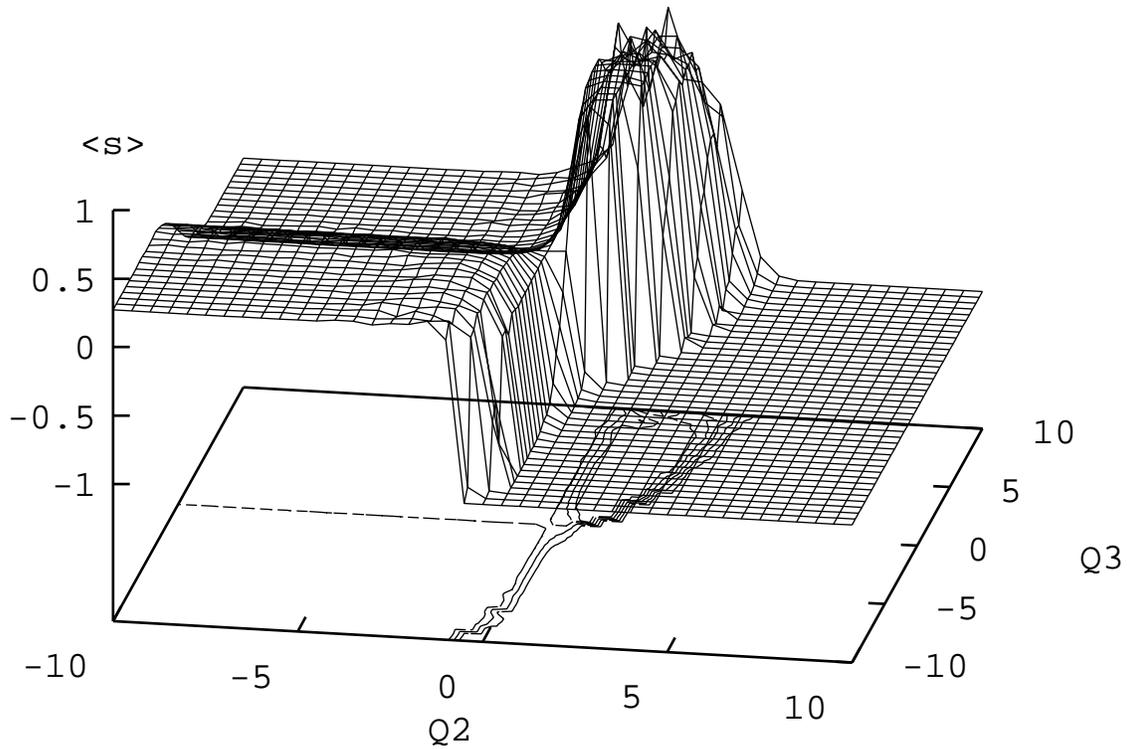

Fig. 4: Spin expectation value as a function of the 2- and 3-spin couplings $Q_2$ and $Q_3$, respectively.

The critical exponent of the honeycomb lattice is given by $e^{2K_c} = 2 + \sqrt{3}$ and thus for the Kagomé lattice

$$e^{4Q_c} = 3 + 2\sqrt{3}$$
$$Q_c = -\tfrac{1}{2}(\lambda c_2)_c = 0.4643 \ . \tag{17}$$

The critical coupling $Q_c$ is positive which reflects the fact that the antiferromagnetic Kagomé lattice is always frustrated even at zero temperature.

Switching on the 3-spin interactions we performed numerical simulations of the system in the coupling plane $(Q_2 = -\tfrac{1}{2}\lambda c_2, Q_3 = -\tfrac{1}{3}\lambda c_3)$. Figure 4 shows the spin expectation value in a certain range of coupling constants. For $Q_3 = 0$ the conventional Ising model on a Kagomé lattice is recovered and for $Q_3 \neq 0$, $Q_2 < 0$ the 3-spin coupling removes the frustration leading to ordered phases: Depending on sign($Q_3$) the spins denoted in Figure 3 by open dots take the values $\sigma = \pm 1$ whereas the full dots take $\sigma = \mp 1$ giving rise to an expectation value $\langle\sigma\rangle = \pm\tfrac{1}{3}$.

We now return to the application of the spin system to the gravitational system. To investigate different types of measures the external field term has to be taken into account. Further the couplings depend on the value of $\epsilon$ according to (8). Figure 5 depicts $\langle\sigma\rangle$ versus $\epsilon$ for the uniform and the scale invariant measure and for the measure leading to a cancellation of the term linear in $\sigma$. The upper curve can be read off directly from Figure 4 and represents only a small detail near the origin. The 3-spin coupling is in the entire range of $\epsilon$ not strong enough to remove the frustration and therefore $\langle\sigma\rangle \approx 0$. Both other curves show a favored occurence of negative spins, corresponding via $q_l = 1 + \epsilon\sigma_l$ to short links reflecting the expected tendency of the lattice to shrink [2].

The motivation for this exploratory study was to approximate quantum gravity by a spin system. It is straightforward to extend this approach from the trivial two-dimensional Regge-Einstein action to scalar fields or physical spin fields. To allow for more than two link lengths an extension to $Z_N$ is possible leading to an increase in the number of coefficients $c_i$. Recently, in three dimensions the Regge-Ponzano model has been considered as a q-deformed su(2) spin network [5]. A generalization of our approach to higher dimensions is possible in principle but becomes more complicated. Notice that in two dimensions only the three links of a triangle are coupled in the action. In three dimensions one is faced with terms of $6^{th}$ order at least, and with additional contributions from the Regge action. The situation is even more complicated in four dimensions where one has to deal with 10 links in each simplex.

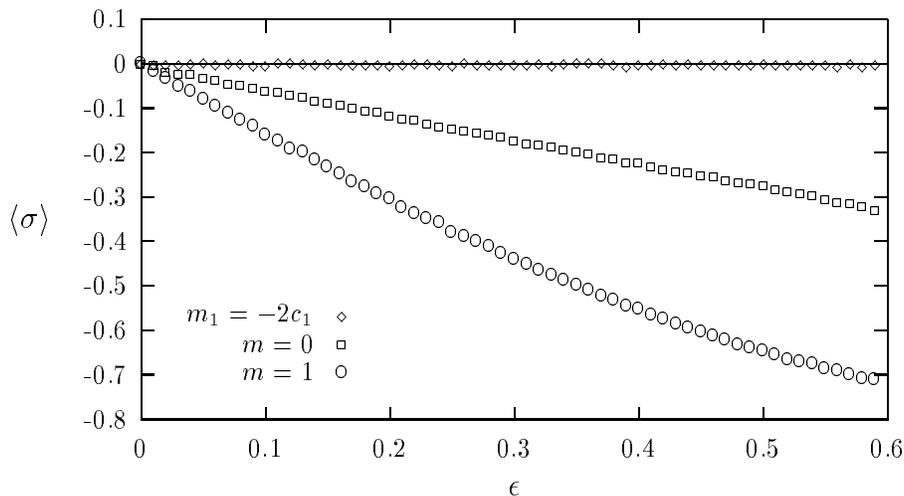

Fig. 5: $\langle \sigma \rangle$ versus $\epsilon$ for $m_1 = -2c_1$ (no external field), $m = 0$ (uniform measure) and $m = 1$ (scale invariant measure). In all cases the cosmological constant is $\lambda = 1$.